\documentclass[aps,pra,10pt,twocolumn,superscriptaddress]{revtex4}
\usepackage{amsmath}
\usepackage{amssymb}
\usepackage{bbm}
\usepackage{graphicx}
\usepackage{framed}
\usepackage{color}
\usepackage{hyperref}
\usepackage{epstopdf}
\usepackage{dsfont}
\usepackage{amssymb}
\usepackage{amsmath}
\usepackage{amsthm}
\usepackage{wrapfig}
\usepackage{relsize}
\usepackage{bm}
\usepackage{textcomp}

\newlength{\eqboxstorage}

\begin{document}

\title{Slow dynamics and thermodynamics of open quantum systems}

\author{Vasco Cavina}
\affiliation{NEST, Scuola Normale Superiore and Istituto Nanoscienze-CNR, I-56126 Pisa, Italy}
\author{Andrea Mari}
\affiliation{NEST, Scuola Normale Superiore and Istituto Nanoscienze-CNR, I-56126 Pisa, Italy}

\author{Vittorio Giovannetti}
\affiliation{NEST, Scuola Normale Superiore and Istituto Nanoscienze-CNR, I-56126 Pisa, Italy}

\begin{abstract}
We develop a perturbation theory of quantum (and classical) master equations with slowly varying parameters, applicable to systems which are externally controlled on a time scale much longer than their characteristic relaxation time. 
We apply this technique to the analysis of finite-time isothermal processes in which, differently from quasi-static transformations, the state of the system is not able to continuously relax to the equilibrium ensemble.
Our approach allows to formally evaluate perturbations up to arbitrary order to the work and heat exchange associated to an arbitrary process. Within first order in the perturbation expansion, we identify a general formula for the efficiency at maximum power of a finite-time Carnot engine. We also clarify under which assumptions and in which limit one can recover previous phenomenological results as, for example, the Curzon-Ahlborn efficiency. 
\end{abstract}
\maketitle

A central result in the study of open quantum systems~\cite{Breuer2002, Rivas2012} is the Markovian Master Equation (MME) approach which, 
under realistic assumptions, describes the temporal evolution of a system of interest $A$ induced by a weak coupling with 
a  large external environment $E$. 
This consists in a first order linear differential equation  
$\dot{\rho}(t)=\mathcal L [{\rho}(t)]$ where $\rho(t)$ is the density matrix of $A$ and 
 where the generator of the dynamics is provided by a quantum Liouvillian  superoperator $\mathcal L$ that can be casted in the so called Gorini-Kossakowski-Sudarshan-Lindblad form~\cite{KOS,LIN,GO}.
For autonomous systems the latter does not exhibit an explicit time dependence and the dynamics of $A$ exponentially relaxes to a (typically unique) equilibrium steady state $\rho_0$  identified by the null eigenvector equation 
 $\mathcal L [{\rho}_0]=0$.
MMEs can also be employed to  describe the temporal evolution of $A$ when it is tampered by the presence of slow varying, external driving forces. 
Indeed, as long as these  
  operate on  a time scale  which is much larger than the characteristic bath correlations times
and  the inverse frequencies of the system of interest,
the effective coupling between $A$ and $E$ adapts instantaneously to the driving control, resulting on a MME 
governed by  a time-dependent Liouvillian generator, i.e.  
\begin{equation}\label{Qmaster}
\dot{\rho}(t)=\mathcal L_t[ {\rho}(t)].
\end{equation}
An explicit integration of this equation  is in general difficult to obtain. Yet, if the control forces are  so slow that their associated time scale
 is also larger with respect to the relaxation time of the system induced by the interaction with $E$, 
one expects $\rho(t)$ to  approximately follow the  instantaneous equilibrium state $\rho_0(t)$ that nullifies  ${\mathcal L}_t$. Our aim is to estimate quantitative deviations from  this ultra-slow driving regime. For this purpose we develop a perturbation theory valid in the limit of slowly varying Liouvillians ${\cal L}_t$  and derive a formal solution of Eq.~(\ref{Qmaster}) which allows one to evaluate non-equilibrium corrections up to arbitrary order. 

The main motivation of our analysis is to model
thermodynamic processes and cycles beyond the usual reversible limit which is strictly valid only for infinitely long quasi-static transformations. Finite-time thermodynamics \cite{Andresen2011,Benenti2016} is a well established research field which is focused on this issue and in particular on the tradeoff between efficiency and power of realistic heat engines. Several results in this context have been derived from the geometrical notion of thermodynamic length \cite{Salamon1983},  from non-equilibrium identities known as fluctuation theorems \cite{Campisi2011,Cavina2016}, or from phenomenological models of heat engines \cite{Benenti2016, Curzon1975}. The latter approach led to the identification of a quite general value for the efficiency at maximum power which is the celebrated Curzon-Ahlborn (Chambadal-Novikov) efficiency \cite{Curzon1975,Chambadal1957,Novikov1957}.  
Our framework  is complementary to previous approaches since it allows to explicitly express irreversible thermodynamic quantities ({e.g.} heat and work) in terms of the Liouvillian operator that governs the system dynamics. In this way we identify a general link between the frequency scaling of the spectral density and the efficiency of finite-time Carnot heat engines, clarifying for which kind of thermal baths the Curzon-Ahlborn result or other particular limits can be recovered.

Similar questions and problems have been addressed in the literature with different aims and methods.  Finite-time quantum thermodynamics \cite{Alicki1979, Kosloff2013,Brandner2016} and Brownian quantum engines  \cite{Schmiedl2007,Dechant2016} were studied using the formalism of open quantum systems. In particular single-qubit heat engines subject to Markovian dissipation were considered in \cite{Szczygielski2013,Esposito2009, Esposito2010bis,Wu2006}. The impact of the bath spectral density on the efficiency of quantum engines was also noticed in the context of autonomous heat pumps \cite{Correa2014,Palao2016} and single-qubit minimal machines \cite{Gelbwaser2013}.  
Universal features and bounds for  the efficiency at maximum power of finite-time Carnot cycles  were identified in \cite{Esposito2010,Schmiedl2007,Vandenbroeck2005,Johal2016} for generic heat engines. Other results were obtained combining MMEs and linear response theory \cite{Davies1978,Bonanca2014,Benenti2016} and similar approaches were used to demonstrate the universality of heat engines in the limit of infinitesimal cycles \cite{Uzdin2015}.
Outside the field of thermodynamics, our theory of slowly driven open quantum systems also contributes to the current research activity on quantum adiabatic driving. Several generalizations of the adiabatic theorem to open quantum systems have been already proposed \cite{Davies1978, Carollo2006, Oreshkov2010, Albash2012}. Our aims and results are however different: we are not interested in the derivation of a modified {MME} for time dependent Hamiltonians but only on the dynamical evolution of $A$ for an assigned time dependent Liouvillian ${\cal L}_t$ that, for all $t$,  admits a  unique (instantaneous) equilibrium state. This kind of adiabatic approach has recently proved to be effective in the study of general time dependent Liouvillians with higher dimensional kernels \cite{Avron2012,Venuti2016,Albert2016} .
We finally would like to stress that, by replacing ${\rho}(t)$ with a probability {vector}
all the results that we are going to present in the following are  directly applicable also to classical continuous-time Markov processes \cite{Norris1997}. \\
{\it Slow Driving Perturbation theory.--- } 
Let us consider the case of an open quantum system $A$ evolving as in Eq.~(\ref{Qmaster}) under the action of a quantum Liuovillian ${\cal L}_t$ which exhibits an explicit temporal dependence induced by the external modulation of some control parameters, say the value of a magnetic field or the intensity of a laser which are gradually changed according to 
some assigned protocol. In what follows we shall assume that for all $t$, ${\cal L}_t$ admits a unique zero (instantaneous) eigenstate $\rho_0(t)$ and that all the other eigenvalues have a strictly negative real part (in this case the map  is said to be relaxing or mixing \cite{Rivas2012, Burgarth2013}).
This causes the system to exponentially converge to the instantaneous steady state, for a fixed value of the external modulation:

\begin{align} \label{ASS}
{\it  Assumption \; 1:}   \phantom{xxxx} 
  &\lim_{s\rightarrow \infty}e^{{\mathcal L}_t s} [\rho]= \rho_0(t), \;\: \forall \rho , t.    \hspace{0.5cm}
\end{align} 
Under these conditions, one can easily verify that for an infinitely slow modulation of ${\cal L}_t$, the system $A$ will be forced to follow
 quasi-statically the trajectory determined by the time-dependent density matrix $\rho_0(t)$, i.e. 
 \begin{eqnarray} 
 \rho(t) \simeq \rho_0(t)\;. \label{ULTRA} 
\end{eqnarray} 
Physically this follows from the fact that, in this regime,  there is enough time  for $A$ to track the instantaneous equilibrium states defined by ${\cal L}_t$. This solution well approximates the dynamics of  realistic configurations  where, for instance, the system of interest is in thermal contact with a reservoir while being subject to a quasi-static external control, continuously relaxing to the instantaneous Gibbs ensemble. Notice however that at this stage of the analysis $\rho_0(t)$ can be an arbitrary quantum state, covering more general open evolutions including also systems in contact with engineered non-thermal baths (e.g. squeezed environments, negative temperatures, artificial dissipative maps, {etc.}). To characterize  deviations from the quasi-static solution Eq.~(\ref{ULTRA}) we find  convenient to 
introduce the following time-rescaled quantities
\begin{eqnarray}
\tilde{\mathcal L}_{t'} &= \mathcal L_{\tau t'},  \qquad 
\tilde \rho(t') &= \rho(\tau t'), 
\end{eqnarray} 
where $\tau$ is the duration of the protocol, i.e. the  total time interval on which the system evolves under the influence of the external control.
With this choice Eq.~(\ref{Qmaster})  can be expressed as 
\begin{align}\label{tildemaster}
 \dot {\tilde \rho}(t')= \tau \tilde{\mathcal L}_{t'} [\tilde \rho(t')],
\end{align} 
the dynamics being confined now into the unit interval $t'\in[0,1]$. In this way the total duration of the protocol appears only as a multiplicative factor while all the information about its ``shape" is contained in $\tilde{\mathcal L}_{t'}$.
Notice also that while $\tilde{\mathcal L}_{t'}$ is independent of $\tau$, the time-rescaled solution $\tilde \rho(t')$ of Eq. \eqref{tildemaster} is not. In particular
 the quasi-static solution~(\ref{ULTRA}) can be recovered from Eq.~(\ref{tildemaster}) in the asymptotic limit $\tau \rightarrow \infty$.  Therefore we look for a perturbation expansion of the solution of Eq.~(\ref{tildemaster})  in powers of $1/\tau$:
 \begin{align}\label{series}
 \tilde \rho(t')= \tilde \rho_0(t') + \frac{\tilde \rho_1 (t')}{\tau}+ \frac{\tilde \rho_2 (t')}{\tau^2}+\dots .
 \end{align}
where normalization implies that all perturbations are traceless
\begin{align} \label{traceless}
{\rm tr}[\tilde \rho_j (t') ]=0, \quad \forall j>0.
\end{align}
A rigorous mathematical analysis of the convergence properties of the series \eqref{series}  is beyond the aim of this work. For our purposes it is sufficient that Eq. \eqref{series}, truncated up to a finite order, provides a good 
approximation of the dynamics. The validity of this approach is verified in several numerical examples presented in the Supplemental Material (Appendix A,B and C). 

Substituting Eq.\ \eqref{series} in both sides of Eq.\ \eqref{tildemaster}, and equating the terms proportional to the same powers of $1/\tau$, we get the following set of recursive relations:
\begin{align}
0 &= \tilde {\mathcal L}_{t'}[ \tilde \rho_0(t')] , \label{ric0} \\
\dot{\tilde \rho}_{j}(t') &= \tilde {\mathcal L}_{t'} [ \tilde \rho_{j+1}(t')],\ \quad j=0,1, \dots\, . \label{ricj}
\end{align}
As expected, Eq.\ \eqref{ric0} implies that $\tilde\rho_0(t')$ is the (time-rescaled) instantaneous steady state of the Liouvillian. 
Moreover from Eq.\ \eqref{ricj}, all finite-time perturbations can be recursively obtained. Indeed, exploiting Eq. \eqref{traceless}, Eq.\ \eqref{ricj}  can be univocally inverted
by introducing the projector $\mathcal P$ on the traceless subspace, i.e.   $\mathcal P[X]=   X- {\rm tr}(X)\mathbb{I}/d$ with $\mathbb{I}$ being the identity operator on $A$
and $d$ the dimension of its Hilbert space. Now since ${\rm tr} [\rho_0(t)]=1\neq 0$, the Liouvillian within the subspace of traceless operators is invertible and we also have $(\mathcal P \mathcal L_t \mathcal P )^{-1}=(\mathcal L_t \mathcal P )^{-1}$.
Accordingly for all $j$ we can write Eq. \eqref{ricj} as $\tilde{\rho}_{j+1}(t')=  (\tilde{\mathcal L}_{t'} \mathcal P)^{-1} \dot{\tilde{\rho}}_{j}(t')$, 
which by iteration yields explicit formulas for each perturbative term:
\begin{align} \label{rhotp}
 \tilde \rho_j(t')= \left[(\tilde{\mathcal L}_{t'} \mathcal P)^{-1} \frac{d}{dt'} \right]^j \tilde{\rho}_0(t').
 \end{align}
Switching from time-rescaled variables back to the original ones, the full solution of the MME can be compactly expressed as an operator geometric series:
\begin{align}\label{rhocompactnormal}
\rho(t) = \frac{1}{1- ({\mathcal L}_{t} \mathcal P)^{-1}\frac{d}{dt}}  \rho_0(t).
\end{align}
It is worth  stressing that the above solution is unique and independent on the initial conditions. This is due to the fact that, at this level of approximation, we are neglecting any term exponentially decaying in $\tau$. In other words, the perturbations that we are considering are intrinsic to the Liouvillian operator without any influence from the initial state. In practice this means that all exact solutions of the master equation, after an exponentially short transient depending on the initial conditions,  will converge to the asymptotic solution \eqref{rhocompactnormal}. This is a fundamental difference with respect to the Hamiltonian adiabatic theorem in which finite-time corrections depend on initial conditions and decay exponentially in $\tau$. 
  In the next sections we present  a couple of relevant applications of the results presented above to the context of finite-time thermodynamics.   

{\it Finite-time isothermal process.--- } 
Let us consider the case of a thermal MMEs~(\ref{Qmaster}) describing the 
 dynamical evolution of $A$ induced by an external driving that slowly modifies its Hamiltonian $H(t)$ while the system is
  constantly kept in thermal contact 
 with a bath of fixed inverse temperature $\beta$.  
  In this scenario the instantaneous equilibrium state  of the problem can be identified with the Gibbs density matrix
$\rho_0(t)=\exp[-\beta H(t)]/ Z (t)$, with $Z(t)= {\rm tr}\{\exp[-\beta H(t)]\}$ being the associated partition function. Exploiting the derivation of the previous section
we wish to determine how departures from the associated quasi-static trajectory~(\ref{ULTRA}) influence 
the thermodynamic properties of the process. For this purpose we remind that the mean energy and the von Neumann  entropy of $A$ are given by  $U(t)={\rm tr}[H(t)\rho(t)]$ and $S(t)= - {\rm tr}[\rho(t) {\rm log}(\rho(t))]$, respectively. Moreover, following a common approach \cite{Alicki1979,Kosloff2013,Anders2013} we identify the mean heat absorbed by the system  during the time interval $[0,\tau]$ with
 \begin{align}\label{heat} 
 Q = \int_0^{\tau} {\rm Tr}[\dot{\rho}(t) H(t)] dt =  \int_0^{1} {\rm Tr}[\dot{\tilde{\rho}}(t') \tilde{H}(t')] dt' ,
\end{align}
and the mean work done on the system with
\begin{align}\label{work} 
W = \int_0^{\tau} {\rm Tr}[ \rho(t) \dot{H}(t)] dt=  \int_0^{1} {\rm Tr}[ {\tilde \rho}(t') \dot{\tilde{H}}(t')] dt',
\end{align} 
such that the first law of thermodynamics is obtained as $\Delta U =U(\tau)-U(0)= W+Q$. Now, since the time-rescaled Hamiltonian $\tilde H(t')$ does not depend on $\tau$ but only on the shape of the protocol, all the previous  quantities are influenced by $\tau$ only through the solution of the master equation $\rho(t)$ for which we know how to evaluate each perturbative term of the series \eqref{series}. Therefore we can write $X= X_0 + X_1/\tau+X_2/\tau^2+\dots$  with $X= U,S,W,Q$, and we can easily evaluate each term using Eq.\ \eqref{rhotp}. 
For example, at the zeroth order approximation we recover the standard results of equilibrium thermodynamics:
$U_0(t) = - \frac{\partial}{\partial \beta} \log Z(t)$, $S_0(t) =(1- \frac{\partial}{\partial \beta})  \log Z(t) $, $Q_0 = \Delta S_0 /\beta$ and $W_0 = \Delta U_0- \Delta S_0 / \beta$.
The corresponding first order irreversible corrections are instead
\begin{align} \label{termoquant1}
U_1(t)&= {\rm tr} \left[\tilde H(t')   \tilde{\rho}_1(t')  \right] _{t'=t/\tau} , \\
S_1(t)&= -{\rm Tr}[ \tilde \rho_1(t') \log( \tilde \rho_0(t'))]_{t'=t/\tau} =  \beta U_1(t) , \\
Q_1& =\int_0^1{\rm tr} \left[ \tilde{H}(t')  \dot {\tilde \rho}_1(t')  \right ] dt' 
\label{Q1} \\
W_1 &=  \Delta U_1-Q_1,
\end{align}
with $\tilde{\rho}_1(t)$ as in (\ref{rhotp}).
It is worth noticing that, independently from the  selected form of the MME the first law of thermodynamics is valid at the level of each perturbative coefficient $\Delta U_j= W_j+Q_j$ and that, for an initial Gibbs state $\rho(0)=\rho_0(0)$, the second law can be expressed as $Q \le Q_0$ or equivalently as $W\ge W_0$. Taking the limit of large $\tau$, this implies that  
$Q_1 \le 0$ and $W_1\ge 0$.  Moreover, if we consider the time-reversed process $\tilde {\mathcal L}^{\leftarrow}_{t'}= \tilde {\mathcal L}_{1-t'}$,
then it is easy to check that odd-order perturbations are invariant while even-ored perturbations change sign  $Q_{j}^{\leftarrow}= (-1)^{j+1} Q_{j}$ and  $W_{j}^{\leftarrow}= (-1)^{j+1} W_{j}$.\\
{\it Finite-time Carnot cycles.--- } As a second application of our slow driving perturbative approach consider the case where 
$A$, initialized in a Gibbs state with Hamiltonian $H_A$, evolves following a Carnot cycle composed by: 

\begin{enumerate}
\item Isothermal expansion: the system is put in contact with a hot bath of temperature $T_H$ and the Hamiltonian is slowly changed from $H^{A}$ to $H^{B}$, in a time interval $\tau_H$.
\item Adiabatic expansion: the Hamiltonian is suddenly changed from $H^B$ to $(T_C /T_H) H^B$.
\item Isothermal compression: the system is put in contact with a cold bath of temperature $T_C$ and the Hamiltonian is slowly changed from $(T_C /T_H)  H_B$ to $(T_C /T_H) H_A $, in a time interval $\tau_C$.
\item Adiabatic compression: the Hamiltonian is suddenly changed from $(T_C /T_H) H_A$ back to $H_A$.
\end{enumerate}

\noindent The scaling factor $T_C /T_H $ characterizing the adiabatic operations is chosen to ensure that the stationary (Gibbs) state $\rho_0(t)$ evolves continuously in time during the cycle such that, for infinitely long processes, the system remains always in equilibrium without irreversible jumps. In addition to this standard requirement we also assume that $\rho_0(t)$ is sufficiently smooth, so that all the derivatives appearing in our perturbation theory are well defined along the cycle. Specifically, since in what follows we are going to consider only first order perturbations, we assume: \\

\noindent {\it Assumption 2:  $\rho_0(t)$ is continuous and differentiable. }\\

\noindent We also assume that, apart from the Hamiltonian scaling factor $T_C /T_H $ and its time length, the cold isothermal protocol is the time-reversal of the hot one: $\tilde{H}^{C}(t')=  (T_C /T_H)  \tilde{H}^{H}(1-t') $. In terms of the associated time-rescaled Gibbs states, this is equivalent to: 
\begin{align} \label{timerev}
\it  \hspace{-0.3  cm} Assumption\;  3: \qquad &  \tilde{\rho}^{C}(t')= \tilde{\rho}^{H}(1-t') \;.  \hspace{1.5 cm}
\end{align} 
The latter assumption is common in many realistic heat engines and can be relaxed if we are free to optimize the shape of the two isothermal processes. Indeed in this case, within first order in the perturbation expansion that we are going to present later, the maximum output power is obtained when the driving protocol respects the time-reversal symmetry condition  \eqref{timerev}.

Now we are ready to analyze the performances of our finite-time Carnot engine. In the limit of many cycles the system evolution becomes periodic ($\Delta U=0$) and the work per cycle depends only on the heat exchanged in the hot and cold isothermal processes:  $W= -Q^H - Q^C$. 
The power is the ratio between the work extracted and the time length of the cycle $P=-W/(\tau_H+\tau_C)=(Q^H+Q^C)/(\tau_H+\tau_C)$,
while the thermodynamic efficiency is defined as $\eta=-W/Q^H= 1+ Q^C/Q^H$.
If we substitute the perturbative expansion, keeping only first order terms in $\tau_H$ and $\tau_C$, we obtain
\begin{align}
P& \simeq \frac{ Q_0^H+ Q_1^H / \tau_H  + Q^C_0 +Q^C_1/\tau_C }{ \tau_H + \tau_C} , \label{Ptau} \\
\eta& \simeq 1 +  \frac{ Q^C_0 + Q^C_1/ \tau_C } {Q_0^H+Q_1^H /\tau_H}. \label{efftau} 
\end{align}
In the quasi-static limit $\tau_C,\tau_H \rightarrow \infty$,  the power tends to zero and the efficiency reaches the Carnot limit
\begin{align}
\eta_{\rm C}= 1 + Q^C_0  / Q_0^H= 1- T_C/T_H. \label{Carnot}  
\end{align}
The power can be maximized with respect to $\tau_H$ and $\tau_C$, taking into account the physical constraints $Q_0^H>0$, $Q_0^C,Q_1^H , Q_1^C<0$. The corresponding efficiency at maximum power can be easily computed  (see {\it e.g.} \cite{Schmiedl2007}):
\begin{align} \label{effMP}
\eta^*=\left[  \frac{2}{\eta_{\rm C}} - \frac{1}{1 +  \sqrt{Q_1^C/Q_1^H}}  \right]^{-1} . 
 \end{align}
 We also notice that, in the particular case in which heat corrections are proportional to the temperature,  Eq. \eqref{effMP} reduces to the efficiency derived in Ref. \cite{Esposito2010}.
 
\begin{figure}[t!]
\includegraphics[scale=.37]{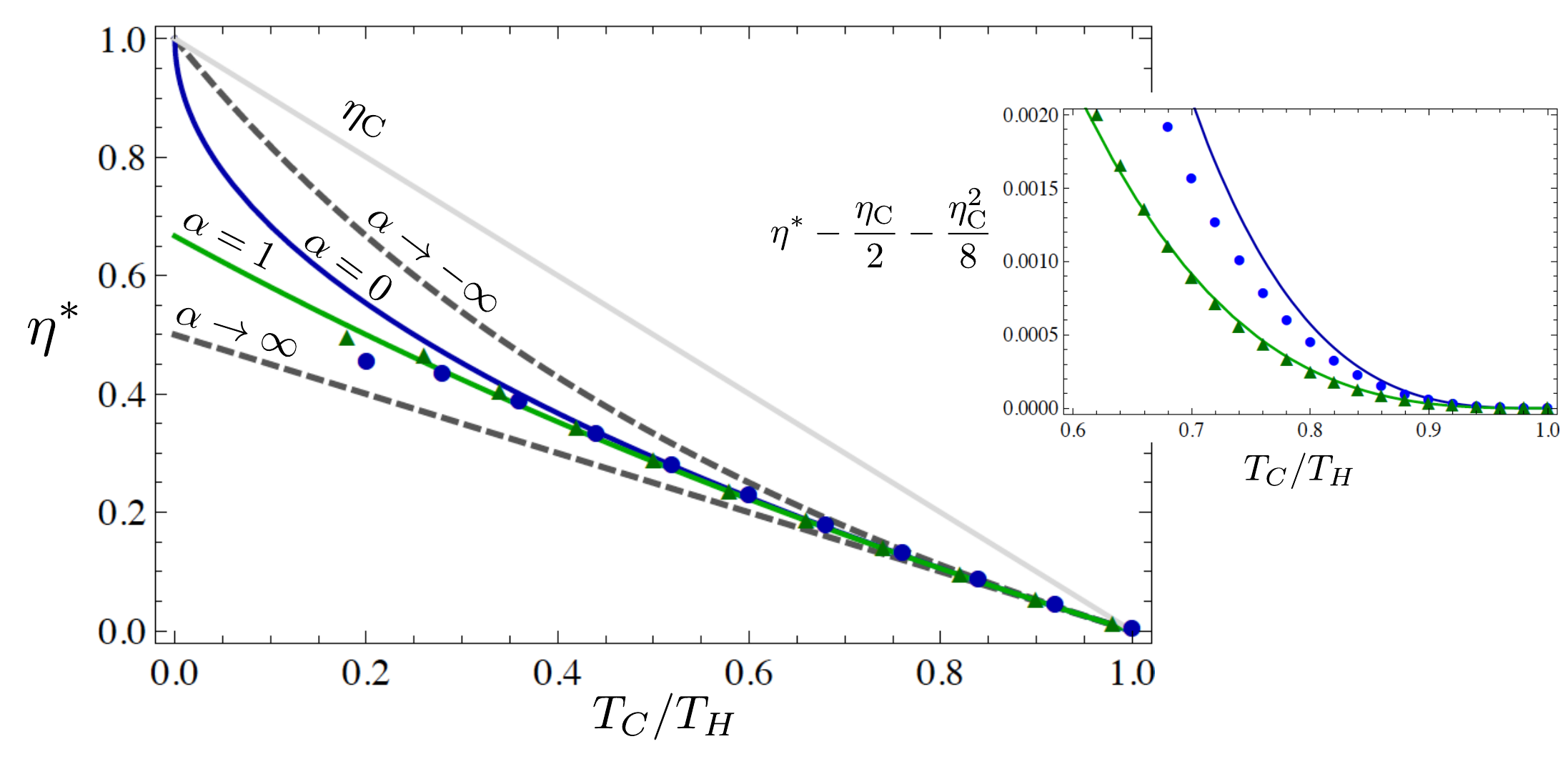}
\caption{Efficiency at maximum power $\eta^*$ with respect to the temperature ratio $T_C/T_H$  for different values of the spectral density exponent $\alpha$, see Eq.~(\ref{effalpha}). Light gray line represents the Carnot bound; blue circles and green triangles are  exact numerical results based on a two-level system heat engine coupled to a flat ($\alpha=0$) and Ohmic bath ($\alpha=1$) respectively. The details of the simulation are reported in the Supplemental Material (Appendix C).}\label{fig1}
\end{figure}

We can now make use of Eq.\ \eqref{Q1} and \eqref{rhotp} to express both $Q_1^H$ and $Q_1^C$ in terms of the Liouvillians $\mathcal L_{t'}^H$ and $\mathcal L_{t'}^C$ describing the two isothermal processes. Both irreversible heat corrections depend on the particular choice of the protocol however we find that their ratio  has the following scaling:
\begin{align} \label{QT}
Q_1^C/Q_1^H = (T_C/T_H)^{1-\alpha},
\end{align}
where $\alpha$ is the frequency exponent of the bath spectral density $J(\omega)\propto \omega^{\alpha}$ which is assumed to be the same for both the hot and the cold reservoirs.  The proof of Eq.\ \eqref{QT} can be found in the Supplemental Material (Appendix D) and follows from the time-reversal condition \eqref{timerev} and a general scaling property characterizing all MME obtained from standard microscopic models. 
Substituting \eqref{QT} in \eqref{effMP}, we get our final result (see Fig.\ \ref{fig1} for a plot):
\begin{align} \label{effalpha}
\eta^*=\left[  \frac{2}{1-T_C/T_H} - \frac{1}{1 +  (T_C/T_H)^{(1-\alpha)/2}}  \right]^{-1}.
 \end{align}

The peculiarity of the universal expression \eqref{effalpha} is that, by changing the parameter $\alpha$, it interpolates between several results which were previously obtained in the literature. For example, for a flat spectral density ($\alpha=0$) we recover the Curzon-Ahlborn \cite{Novikov1957, Chambadal1957, Curzon1975} efficiency  $\eta^* |_{\alpha=0}=1- \sqrt{ T_C/T_H}$, while for an Ohmic bath $\alpha=1$ we get $\eta^*|_{\alpha=1}=2 \eta_{\rm C}/ (4-\eta_{\rm C})$ which is the efficiency obtained by Schmiedl and Seifert for a specific Brownian engine \cite{Schmiedl2007}. Finally, consistently with known lower and upper bounds \cite{Schmiedl2007, Esposito2010, Benenti2016},   by taking the two limits of infinitely super-Ohmic or sub-Ohmic bath, we get  $\eta^*|_{\alpha \rightarrow \infty}=\eta_{\rm C} /2  \le \eta^*  \le \eta_{\rm C}/ (2-\eta_{\rm C})=\eta^*|_{\alpha\rightarrow -\infty}$. \\
In Ref.s \cite{Gelbwaser2013,Correa2014,Palao2016}, the efficiency of minimal models of heat engines and refrigerators was also linked to the spectral density. Such models are quite different from the Carnot cycles considered in this work but still one can find a qualitative agreement with our results. \\
A final remark should be made on the range of applicability of Eq. \eqref{effalpha}. Its derivation follows from the first order expansion performed in Eq.s  \eqref{Ptau} and \eqref{efftau} which is not guaranteed to be always valid in the regime of maximum power, especially for $T_C/T_H  \simeq 0$ where the optimal values of $\tau_H$ and $\tau_C$ could become too small. This is confirmed by the single-qubit exact simulation shown in Fig.\ \ref{fig1} where we observe   deviations from \eqref{effalpha} for small values of $T_C/T_H$. It is interesting then to expand Eq. \eqref{effalpha} around the opposite regime, i.e. for small values of  $\eta_{\rm C}=1-T_C/T_H$, obtaining 
\begin{align} \label{effExp}
\eta^*= \eta_{\rm C}/2 + {\eta_{\rm C}}^2/8 + {\eta_{\rm C}}^3 (2- \alpha ) / 32 + O ({\eta_{\rm C}}^4).
 \end{align}
We notice that the first and second order coefficients ({\it i.e.} $1$ and $1/8$) are independent of $\alpha$ and correspond to the same values in the Taylor expansion of the Curzon-Ahlborn \cite{Curzon1975,Esposito2009} and of the Schmiedl-Seifert \cite{Schmiedl2007} efficiencies. This also implies that, up to second order in $\eta_{\rm C}$, our results are in agreement with previous analyses based on linear response thermodynamics \cite{Johal2016,Benenti2016}.

{\it Conclusions and outlook}.--- 
We have derived a perturbation theory for the solution of generic master equations with slowly varying coefficients. We focused in particular on finite-time thermodynamic processes beyond the reversible limit. Our analysis allows to analytically derive finite-time thermodynamic quantities in terms of the Liouvillian operator. We have also shown that, for a Carnot cycle, the efficiency at maximum power can be reduced to a universal formula depending only on the temperature ratio and on the scaling of the bath spectral density.  Other implications of the perturbation theory presented in this work could also be studied in the future, in particular within the general context of quantum adiabatic driving. \\

{\it Acknowledgments}.--- We thank L. C. Venuti, M. Fraas, A. Carollo, V. V. Albert, L. A. Correa, D. Alonso and D. Gelbwaser, for fruitful discussions.

 \newpage

  \section*{SUPPLEMENTAL MATERIAL}
  
   \appendix
  \section{Slow driving of a two-level system}
  
  In this section we compare the perturbation theory presented in the main text with the exact open dynamics of a two-level system. Consider a system with Hamiltonian 
  \begin{align}
  H= \frac{\hbar\omega}{2}  \sigma_z + V(t),
   \end{align}  
  where $\omega$ is the energy splitting, $\sigma_z$ is the Pauli matrix and $V(t)$ is a generic perturbation. 
  Moving to interaction picture with respect to $\hbar \omega \sigma_z/2$, we can model the contact of the system with a bosonic heat bath through a standard master equation \cite{Breuer2002A,Rivas2012A}:
  
  \begin{align}
\dot{\rho}(t) &=- \frac{i}{\hbar} \big[V_I(t),\rho(t)]    \nonumber \\
  &+\gamma (N+1) \left(\sigma_- \rho(t) \sigma_+ - \frac{1}{2} \big\{ \sigma_+ \sigma_- ,\rho(t)\big\} \right)  \nonumber \\
  &+\gamma  N \left( \sigma_+ \rho(t)  \sigma_- - \frac{1}{2} \big\{ \sigma_- \sigma_+, \rho(t)\big\}  \right), \label{masterqubit}
  \end{align}
where 
\begin{align}
\gamma &=\gamma_0 \omega^\alpha, \\
N &= [\exp [\beta \hbar \omega]-1]^{-1},
\end{align}
are the damping rate and the mean excitation number respectively, associated to a thermal bath with spectral density $J(\omega)\propto \omega^\alpha$ and inverse temperature $\beta$. 

Now, just to make an example in which off-diagonal coherences are relevant in the dynamics, consider the following perturbation
\begin{align}
V_I= \frac{\hbar \Delta}{2}\, \sigma_x,
\end{align}
and for the moment assume that $\omega$ and $\Delta$ are constant parameters. 
Recasting  the $2\times2$ density matrix into a $4\times1$ complex vector
\begin{equation}
\rho=\left[ \begin{array}{cc}
 \rho_{11} &   \rho_{12}\\
 \rho_{21} &   \rho_{22} 
 \end{array}  \right] \rightarrow  || \rho \rangle=[ \rho_{11},\rho_{12},\rho_{21},\rho_{22}]^{\top},
\end{equation}
it is easy to check that  the projector on the traceless subspace $\mathcal P$ is represented by the matrix 
\begin{equation}
P=\frac{1}{2}\left[
\begin{array}{cccc}
1& 0 & 0 & -1 \\
 0 & 0 & 0 & 0   \\ 
 0 & 0 & 0 & 0   \\
 -1 & 0 & 0 & 1
\end{array}
\right],
\end{equation}
while the Liouvillian superoperator $\mathcal L$ defined by the right-hand-side of Eq. \eqref{masterqubit}  has the following matrix representation:
\begin{equation}\label{Lmatrix}
L=\gamma \left[
\begin{array}{cccc}
 -1-N  & \frac{i \Delta }{2 \gamma} & -\frac{i \Delta }{2 \gamma} & N  \\
 \frac{i \Delta }{2 \gamma} & -\frac{1}{2} (1+2 N)   & 0 & -\frac{i \Delta }{2 \gamma} \\
 -\frac{i \Delta }{2 \gamma} & 0 & -\frac{1}{2} (1+2 N)   & \frac{i \Delta }{2 \gamma} \\
 1+N  & -\frac{i \Delta }{2 \gamma} & \frac{i \Delta }{2 \gamma} & -N 
\end{array}
\right].
\end{equation}
The steady state for which $\mathcal L(\rho_0)=0$ is given by the unique null eigenvector of $L$:
\begin{equation}\label{ssqubit}
||\rho_0 \rangle=z^2
\left[
\begin{array}{c}
N (1+2 N)+(\Delta/\gamma) ^2 \\  -i   \Delta/\gamma \\ i   \Delta/\gamma  \\ (1+N) (1+2 N) +(\Delta/\gamma) ^2
\end{array}
\right],
\end{equation}
with 
\begin{equation}
z^2= \frac{1}{(1+2 N)^2+2 (\Delta /\gamma)^2}.
\end{equation}

Now assume that $\Delta$ is slowly modulated in the time interval $t\in [0,\tau]$. For example, let us take
\begin{align}
\Delta(t)= \Delta_0 \cos[\pi t/\tau ].
\end{align}
In the quasi-static limit $\tau \rightarrow \infty$ we expect the system to follow the instantaneous steady state \eqref{ssqubit}, which now depends on time through the parameter $\Delta(t)$.

\begin{figure}[t!] 
\vspace{1 cm}
\includegraphics[scale=.41]{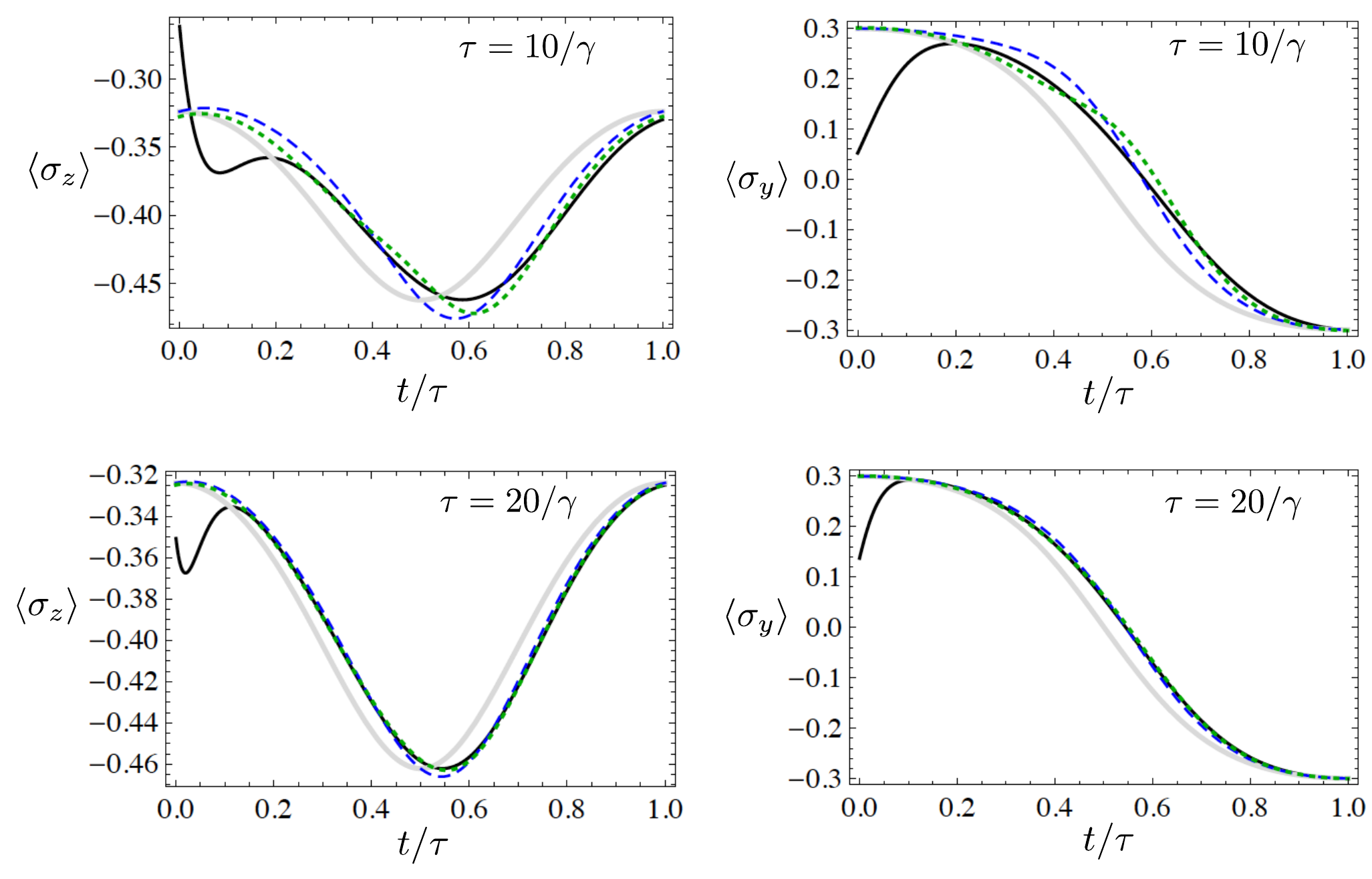}
\caption{Bloch coordinates $\langle \sigma_z(t)\rangle$ (left column) and  $\langle \sigma_y(t)\rangle$ (right column) with respect to the rescaled time $t'=t/\tau$.  Due to the symmetry of the master equation, $\langle \sigma_x(t)\rangle=0$. Exact solution (black line) of the master equation with initial state $\rho(0)=\mathbbm 1/2$, $0$th order approximation (gray line), $1$st order approximation (blue dashed line), $2$nd order approximation (green dotted line).  In the first row we set the time length of the process to $\tau=10/\gamma$, while in the second row the duration is extended to $\tau=20/\gamma$. We observe that the exact dynamics is better approximated by increasing the perturbation order and that, in general, the error reduces for large values of $\tau$. 
 The values of the other parameters are such that $\beta\hbar \omega= \Delta_0/\gamma=1$. }\label{figc}
\end{figure}
For a finite $\tau$ we can apply the perturbation theory presented in the main text. We want to compare the exact numerical solution with the corresponding $0$th, $1$st  and $2$nd order approximations which are given respectively by:

\begin{align}
|| \rho^{(0)}(t) \rangle & =  || \rho_0(t) \rangle  ,  \label{app0} \\
 || \rho^{(1)}(t) \rangle              &= \left [1 + (L(t ) P)^{-1}   \frac{d}{dt}  \right ]  || \rho_0(t) \rangle  ,   \label{app1} \\
  || \rho^{(2)}(t) \rangle              &= \left [ 1 +  (L(t ) P)^{-1}   \frac{d}{dt} +\left[ (L(t ) P)^{-1}   \frac{d}{dt} \right]^2 \right]  || \rho_0(t) \rangle   \label{app2} , 
\end{align}
where the pseudo-inverse of the projected Liouvillian can be explicitly computed.  Omitting the time parameter to simplify the notation, the latter is given by:
{
\begin{align}\label{LPm1}
&(LP)^{-1} = \frac {z^2}{2 \gamma } &\nonumber \\
 &\left[
\begin{array}{cccc}
 - \small{1-2 N} & -\frac{2 i \Delta }{\gamma } & \frac{2 i \Delta }{\gamma } & 1+2 N    \\
 -\frac{2 i \Delta }{\gamma } & -4 \frac{ z^{-2}-(\Delta/\gamma) ^2} {1+2 N} & -\frac{4( \Delta /\gamma)^2}{1+2 N } & \frac{2 i \Delta }{\gamma } \\
 \frac{2 i \Delta }{\gamma } & -\frac{4( \Delta /\gamma)^2}{1+2 N } &-4 \frac{z^{-2}-(\Delta/\gamma) ^2}{1+2 N} & -\frac{2 i \Delta }{\gamma } \\
 1+2N & \frac{2 i \Delta }{\gamma } & -\frac{2 i \Delta }{\gamma } & -1-2 N
\end{array}
\right].
\end{align}
}%

We can finally compare the numerical solution of the master equation \eqref{masterqubit} (for a maximally mixed initial state) with the corresponding analytical approximations (\ref{app0},\ref{app1},\ref{app2}). The results are reported in Fig. \ref{figc} where we observe the, after an initial transient depending on the initial conditions, the dynamics is well captured by the perturbative approximations.

  \section{Finite-time isothermal process of a two-level system }

In this section we explicitly compute the finite time corrections to the heat absorbed by a two-level system during an isothermal process in which the energy splitting $\omega(t)$ is changed in time within the interval $t\in[0,\tau]$.
The system can be described by the previous master equation \eqref{masterqubit} setting $V_I(t)=0$ (the density matrix remains diagonal). In interaction picture the Hamiltonian evolution is canceled and the master equation depends 
on the modulated frequency only via the damping rate and the mean excitation number: 
\begin{align}
\gamma(t) &=\gamma_0 \omega(t)^\alpha, \\
N(t) &= [\exp [\beta \hbar \omega(t)]-1]^{-1}.
\end{align}
Setting $\Delta=0$ in Eq.\ \eqref{Lmatrix}  and dropping the time variable in the parameters $\gamma(t)$ and $N(t)$,  the Liouvillian matrix simplifies to the following block form
\begin{equation}\label{Lmatrix2}
L=\gamma \left[
\begin{array}{cccc}
 -1-N  & 0&0 & N  \\
0& -\frac{1}{2} (1+2 N)   & 0 &0 \\
0& 0 & -\frac{1}{2} (1+2 N)   & 0 \\
 1+N  &0& 0& -N 
\end{array}
\right],
\end{equation}
whose instantaneous eigenvector is the Gibbs state 
\begin{equation}\label{ssqubit2}
||\rho_0(t) \rangle=\frac{1}{2N(t)+1}
\left[
\begin{array}{c}
N(t)	 \\  0  \\ 0 \\ 1+N(t)
\end{array}
\right]= \frac{1}{2} ( || \mathbbm I \rangle + z(t) || \sigma_z \rangle ),
\end{equation}
where 
\begin{align}
z(t)=\frac{-1}{2 N(t)+1}= -\tanh \left[  \frac{\beta  \hbar \omega(t)}{2}\right]
\end{align}
\noindent is the $z$-coordinate of the Gibbs state in the Bloch sphere.
The inverse of the projected Liouvillian \eqref{LPm1} reduces in this case to:
\begin{align}\label{LPm12}
[L(t)P]^{-1} &= \frac {-z(t)}{2 \gamma(t) } 
\left[
\begin{array}{cccc}
 -1 &0& 0& 1    \\
 0& -4 &0& 0\\
 0& 0&-4 & 0\\
1 &0 & 0 &-1
\end{array}
\right],
\end{align}
and it is easy to check that $[L(t)P]^{-1}$ is diagonal in the basis of vectorized Pauli matrices, in particular:
\begin{align}
[L(t)P]^{-1}  || \mathbb I \rangle &=0 .\\
[L(t)P]^{-1}  ||\sigma_z \rangle &=  \frac{z(t)}{\gamma(t)} || \sigma_z \rangle . 
\end{align}
Using the previous equations, the perturbative solution (11) presented in the main text becomes:
\begin{equation}
\rho (t) = \frac{1}{2} \left [ \mathbbm I +    \frac{1}{1-\frac{z(t)}{\gamma(t) } \frac{d}{dt} } z(t)  \sigma_z   \right],
\end{equation}
where we have reintroduced density matrix representation.
In terms of time rescaled variables $\tilde X(t') = X (t' \tau) $, the perturbation terms defined in Eq. (10) of the main text are given by:
\begin{align}
\tilde \rho_j(t')= \frac{1}{2}\left[\frac{\tilde z(t')}{\tilde \gamma(t')} \frac{d}{dt'} \right]^j \tilde z(t') \sigma_z, \quad j>0.
\end{align}
The heat absorbed by the system during the process can be expanded with respect to $1/\tau$ as $Q=\sum_{j=0}^\infty Q_j  \tau^{-j}$, where 

\begin{align}
 Q_j&=  \int_0^1 {\rm tr} [\tilde H(t')  \dot{ \tilde \rho}_j(t') ]dt'\\
       &=\int_0^1 \frac{\hbar \tilde \omega(t')}{2}  \frac{d}{dt'}\left[\frac{\tilde z(t')}{\tilde \gamma(t')} \frac{d}{dt'} \right]^j \tilde z(t') dt' . \label{Qqubitj}
\end{align}
Analogously, for the work done on the system we have:
\begin{align}
 W_j&=  \int_0^1 {\rm tr} [\dot {\tilde H}(t') \tilde \rho_j(t') ]dt'\\
       &=\int_0^1 \frac{\hbar \dot{\tilde \omega}(t')}{2} \left[\frac{\tilde z(t')}{\tilde \gamma(t')} \frac{d}{dt'} \right]^j\tilde z(t') dt' .
\end{align}

 \section{Finite-time Carnot cycle of a two-level system }

In this appendix, we consider a Carnot cycle performed on a two-level system with Hamiltonian $H=\sigma _z \hbar  \omega(t)/2$. The protocol is defined as follow:
\begin{enumerate}
\item Isothermal expansion: the system is put in contact with a hot bath of temperature $T_H$ and the frequency is reduced according to the function $\omega^H(t)=\omega_0 ( \cos(\pi t/\tau_H ) +1)$, in a time interval $t\in [0,\tau_H]$.
\item Adiabatic expansion: the system is decoupled from the bath and the frequency is shifted to  $\omega^C(0)= (T_C /T_H) \omega^H(\tau_H)$. Actually this step is useless in this case since $\omega^H(\tau_H)=0$, but we keep it just for clarity.
\item Isothermal compression: the system is put in contact with a cold bath of temperature $T_C$ and the frequency is increased according to the function $\omega^C(t)= (T_C /T_H)\omega_0 ( \cos[\pi (1-t/\tau_C)] +1)$, in a time interval $t\in [0,\tau_C]$.
\item Adiabatic compression:  the system is decoupled from the bath and the frequency is shifted to back to $\omega^H(0)=(T_C /T_H) \omega^C(\tau_C)$.
\end{enumerate}
Notice that the driving protocol is such that the Bloch coordinate $z(t)=\tanh(\beta(t) \omega(t) /2)$ of the quasi-static Gibbs state $\rho_0(t)$ remains continuous and differentiable during the full cycle. In particular, at the extreme of the two isothermal processes, $\dot \omega(t)=0$ ensuring the continuity of $\dot z(t)$ also at such critical points. Moreover the shape of the hot and of the cold isothermal processes respect the time reversal symmetry, such that 
\begin{equation}\label{trev}
\tilde{z}^C(t')=\tilde{z}^H(1-t').
\end{equation}
Thus we can apply the perturbation theory presented in the main text for estimating the heat exchanged in one cycle. From \eqref{Qqubitj} we have 
\begin{align} 
Q_0^H&= \Delta S /\beta_H, \\
Q_0^C&= -\Delta S /\beta_C, \\
Q_1^H&= {\beta_H}^{\alpha-1}  \int_0^1 F (\tilde z^H(t') ) dt' . \label{intH}\\
Q_1^C&= {\beta_C}^{\alpha-1}  \int_0^1 F (\tilde z^C(1-t') ) dt' .\label{intC}
\end{align}
where the function
\begin{align}
F(\tilde z(t))= -{\rm arctanh}(\tilde z(t)) \frac{d}{dt'} \frac{\tilde z(t') \frac{d}{dt'} \tilde z(t')}{\tilde \gamma_0   [-2 {\rm  arctanh}(\tilde z(t))/\hbar ]^\alpha} ,
 \end{align}
depends on the particular shape of the process only through $\tilde z(t)$. Thus, the time reversal symmetry \eqref{trev} implies that the two integrals in \eqref{intH} and \eqref{intC}
 are equal and we get, consistently with Eq. (23) of the main text,
 \begin{align}\label{ratioqubit}
\frac{Q_1^C}{Q_1^H}=\left(\frac{\beta_C}{\beta_H} \right)^{\alpha-1}=\left(\frac{T_C}{T_H} \right)^{1-\alpha}.
 \end{align}
Optimizing the output power with respect to $\tau_H$ and $\tau_C$ and using Eq.\ \eqref{ratioqubit}, one obtains 
\begin{align} \label{effalphaqubit}
\eta^*=\left[  \frac{2}{1-T_C/T_H} - \frac{1}{1 +  (T_C/T_H)^{(1-\alpha)/2}}  \right]^{-1},
 \end{align}
which is the same formula for the efficiency at maximum power that we derived for general systems in the main text.

\begin{figure}[!]
\includegraphics[scale=.37]{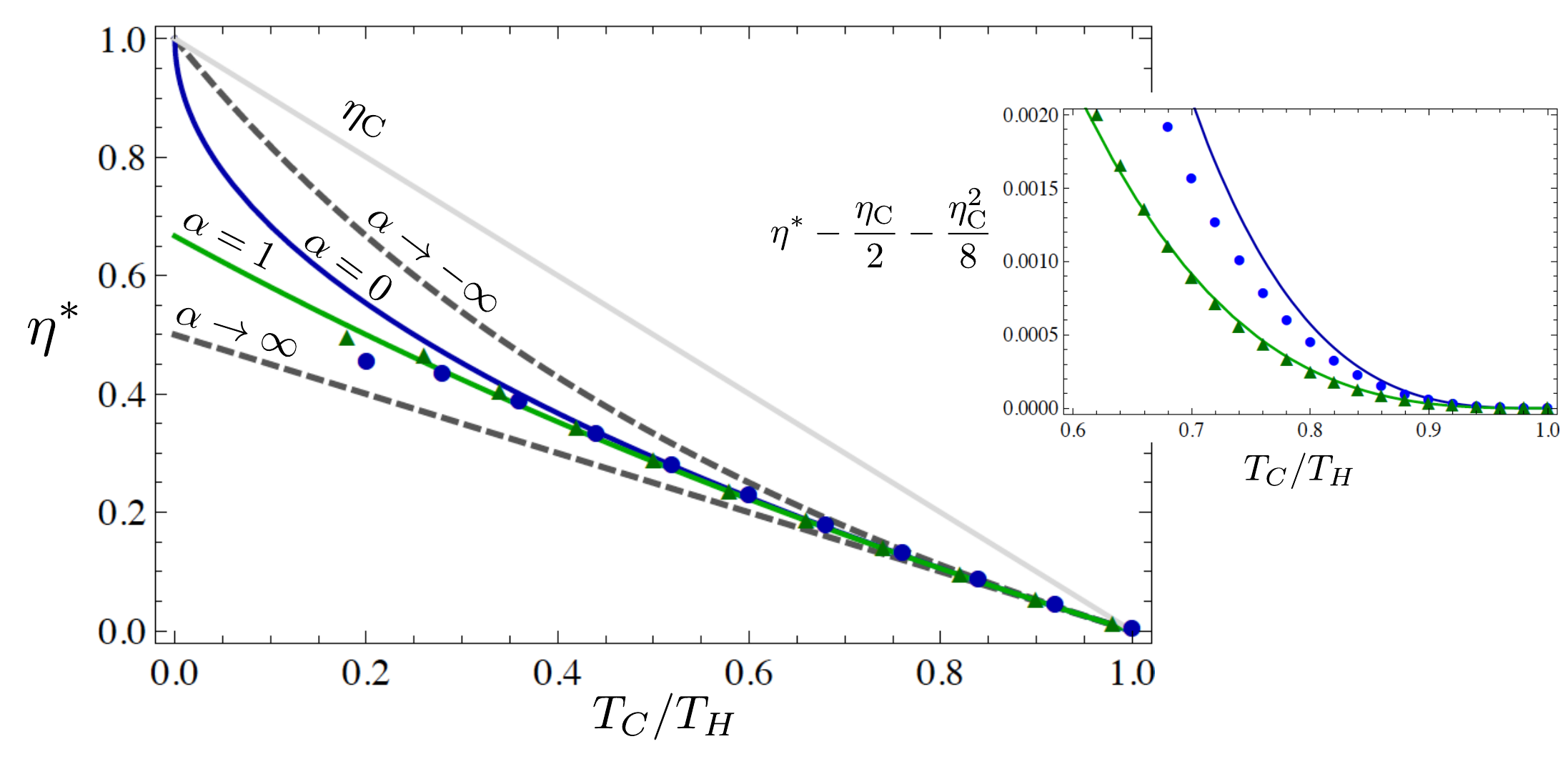}
\caption{Comparison of the analytical estimate and the corresponding exact value of efficiency at maximum power for a Carnot cycle performed on a two-level system.  Blue circles and green triangles are exact numerical results for a flat ($\alpha=0$) and an Ohmic ($\alpha=1$) bath respectively.  Blue and green lines are the corresponding analytical approximations based on Eq.\ \eqref{effalphaqubit}.  The light gray line represents the Carnot bound, while the two dotted lines corresponds to the analytical upper and lower bounds obtained for $\alpha \rightarrow \pm \infty$.  In order to better distinguish the results in the region of large values of $T_C/T_H$, in the inset we plotted the deviation of the efficiency at maximum power with respect to its second order approximation (see Eq.\ (25) of the main text). The other parameters are such that $\beta_H\hbar\omega_0=2$.}  \label{figApp}
\end{figure}

For the same optimal values of $\tau_H$ and $\tau_C$, it is interesting now to compare Eq. \eqref{effalphaqubit} with the corresponding exact result 
\begin{align} 
\eta^{*}_{{\rm exact}}=1+\frac{Q_C}{Q_H},
 \end{align}
 where the heat exchanged in the two processes can be computed from the exact numerical solution of the master equation (evolved for sufficiently many cycles such that any memory of the initial conditions is suppressed).
The comparison between the two values of the efficiency at maximum power is reported in Fig. \ref{figApp}. Notice that exact numerical results are missing for $T_C/T_H$ below a certain threshold. This is due to the fact that for too small values of $T_C/T_H$,  the exact dynamics deviates so much from the idealized Carnot cycle that the system does not produce positive work anymore and cannot be interpreted as an engine. 

 \section{Scaling properties of thermal Liouvillians and heat fluxes}
 
In this final appendix we derive a universal scaling property of thermal Liouvillians which implies the general validity of Eq.\ (23) presented in the main text and already derived, for the particular case of a two-level system, 
in Eq.\ \eqref{ratioqubit}.
Let us start from a typical microscopic model used in the derivation of Markovian master equations \cite{Breuer2002A,Rivas2012A}, in particular we closely follow the continuous bath notation used in \cite{Rivas2012A}. Consider a system of Hamiltonian $H_S$ coupled to a heat bath of harmonic oscillators such that 
the full Hamiltonian can be written as
\begin{align}
H= H_S+ V+ H_B.
\end{align}
where 
\begin{align}
H_B= \int_0^\infty \hbar \omega a_\omega^\dag a_\omega d \omega,
\end{align}
is the Hamiltonian of the bath, $a_\omega$ being the annihilation operator of a mode with frequency $\omega$  obeying the bosonic commutation rule $[a_\omega, a_\omega^\dag]=\delta(\omega-\omega')$.
For the interaction we consider the following potential 
\begin{align}
V=  A \otimes  \int_0^\infty h(\omega) (a_\omega+a_\omega^\dag) d\omega,
\end{align}
where $A$ is an arbitrary system operator while the coupling function $h(\omega)$ simultaneously contains information about the coupling strength and the degeneracy of the bath modes. 
The square of $h(\omega)$ is known as the bath {\it spectral density}, which is usually modeled as a power law (with a suitable cut-off at large frequencies):
\begin{align}
J(\omega)=h^2(\omega)\simeq \frac{\gamma_0}{2 \pi} \omega^\alpha.
\end{align}
The exponent $\alpha$ is a characteristic parameter of the system-bath interaction: {\it e.g.} for $\alpha=0$ the bath is usually called flat, while for $\alpha=1$ the bath is said to be Ohmic.
The operator $A$ can always be decomposed in terms of the eigenoperators of $H_S$:
\begin{align}
 A= \sum_\omega A(\omega)= \sum_{\omega >0}  A(\omega) + A(\omega)^\dag,
\end{align}
where 
\begin{align}
 A(\omega)= \sum_{\substack{{n,m} \\E_m-E_n=\omega}} \langle  n |A |m \rangle  |n \rangle \langle m | .
\end{align}
Moving in interaction picture with respect to $H_S+H_B$, within the usual weak-coupling and secular approximation, one can derive the following Markovian master equation for the reduced state of the system \cite{Breuer2002A,Rivas2012A},
\begin{align}
\dot \rho (t)=\mathcal L [\rho(t)]&=   \sum_{\omega>0} \gamma_0 \omega^\alpha (N(\omega)+1) \mathcal D_{A_\omega}[\rho(t)] +  \nonumber\\
                      &\sum_{\omega>0}  \gamma_0 \omega^\alpha N(\omega) \mathcal D_{A_\omega^\dag}[\rho(t)]. \label{masterHs}
\end{align}
where $D_X(\rho)$ is the Gorini-Kossakowski-Sudarshan-Lindblad dissipator defined as 
\begin{align}
D_X(\rho)= X \rho X^\dag - \frac{1}{2}( X^\dag X \rho + \rho X^\dag X),
\end{align}
and 
\begin{align}
N(\omega)= [ \exp{(\beta \hbar \omega )} -1 ]^{-1}.
\end{align}
One can also show \cite{Breuer2002A,Rivas2012A} that the steady state $\mathcal L [\rho_0]=0$ of the master equation is the Gibbs state $\rho_0 \propto \exp{(-\beta H_S)}$ associated to the system Hamiltonian $H_S$.

Consider now another system-bath configuration in which we perform the rescaling:  $H_S\rightarrow \lambda H_S $ and $\beta \rightarrow  \beta/\lambda$ , for some $\lambda>0$. Clearly the steady state remains the same but what can be said about the master equation?
The quantum Liouvillian defined by the right-hand-side of \eqref{masterHs} is completely determined by the system Hamiltonian $H_S$, the inverse temperature $\beta$ of the bath, the damping constant $\gamma_0$ and the spectral density exponent $\alpha$.
Now, by simply following all the steps of the previous microscopic derivation and observing that the rescaling of $H_S$ implies that all energy gaps change as $\omega \rightarrow \lambda \omega$, it is easy to check that: 
\begin{align}
\mathcal L ( \lambda H_s, \lambda^{-1} \beta,\gamma_0,\alpha)=\mathcal L( H_s,\beta,\lambda^\alpha \gamma_0,\alpha). \label{Lscaling}
\end{align}
Basically, Eq.\ \eqref{Lscaling} states that a rescaling of $H_S$ and of $\beta$ with inverse factors leaves the Liouvillian invariant up to a renormalization of the damping rate.
In interaction picture with respect to $H_S$, one may also rewrite Eq.\ \eqref{Lscaling} simply as 
\begin{align}
\mathcal L ( \lambda H_s, \lambda^{-1} \beta)=\lambda^\alpha \mathcal L( H_s,\beta). \label{Lscaling2}
\end{align}
The validity of Eq.\ \eqref{Lscaling2} also extends to more general interactions of the form $V=\sum_l A_l \otimes B_l$, conditioned on the requirement that all the spectral densities associated to the different bath operators $B_l$ should  scale with the same frequency exponent, {\it i.e.} $J_l(\omega)= \gamma_l \omega ^\alpha/2\pi$.

Now, consider the first order correction to the heat exchanged during an arbitrary isothermal process which, according to the perturbation theory presented in the main text is given by 
\begin{align} \label{Q1general}
Q_1&= \int_0^1{\rm tr} \left[ \tilde{H}(t') \frac{d}{dt'}   (\tilde{\mathcal{L}}_{t'} \mathcal P)^{-1} \frac{d}{dt'}  \tilde \rho_0(t') \right] dt'.
\end{align}
How does the quantity \eqref{Q1general} changes if we preform the rescaling $H\rightarrow \lambda H $ and $\beta \rightarrow \beta/\lambda$ that we have previously discussed? We know that the Gibbs state $\rho_0$
is obviously invariant, moreover exploiting Eq.\eqref{Lscaling2}, we get:
 \begin{align} \label{Q1general2}
Q_1 \xrightarrow{H\rightarrow \lambda H, \: \beta \rightarrow  \beta/\lambda} \;  \lambda^{1-\alpha}Q_1.
\end{align}
We can finally consider the Carnot cycle discussed in the main text. By definition, up to an irrelevant time-inversion, the cold isothermal compression is related to the hot isothermal expansion by exactly the same kind of rescaling that we have just considered above with $\lambda=T_C/T_H$. Therefore from \eqref{Q1general2} we get
 \begin{align} \label{ratiogeneral}
\frac{Q^C_1}{Q^H_1}= \left( \frac{T_C}{T_H} \right)^{1-\alpha},
\end{align}
independently on the particular shape of the driving protocol, corresponding to Eq.\ (23) of the main text.

\end{document}